%% file: main.tex
\documentclass[10pt,journal,compsoc]{IEEEtran}
\usepackage{amsmath,amsfonts}
\usepackage{array}
\usepackage{textcomp}
\usepackage{url}
\usepackage{verbatim}
\usepackage{graphicx}
\usepackage{cite}
\usepackage{algpseudocodex}
\usepackage{tabularx}
\usepackage{float}
\usepackage{color}
\usepackage{xcolor}
\usepackage{soul}
\usepackage{amssymb}
\usepackage{amsthm}
\usepackage{multirow}
\usepackage{booktabs}
\usepackage{dashrule}
\usepackage{arydshln}
\usepackage{enumerate}
\usepackage[linesnumbered,titlenumbered,ruled,vlined,commentsnumbered,noend]{algorithm2e}
\usepackage{catchfilebetweentags}
\usepackage{pifont}
\usepackage{bm}
\usepackage{colortbl}
\usepackage{tcolorbox}
\usepackage{array}
\usepackage{subfigure} 
\graphicspath{{./images/}}
\theoremstyle{definition}
\newtheorem{definition}{Definition}

\newcommand\bs[1]{\boldsymbol{#1}}
\newcommand\nbf[1]{\noindent{\textbf{#1}}}

\newcommand*\revcircled[1]{\tikz[baseline=(char.base)]{\node[shape=circle,draw,fill=black,text=white,inner sep=1pt] (char) {#1};}}
            
\definecolor{mygreen}{HTML}{00c9a7}
\definecolor{1red}{HTML}{ffccd5}
\definecolor{2red}{HTML}{ffb3c1}
\definecolor{3red}{HTML}{ff8fa3}
\definecolor{4red}{HTML}{ff4d6d}
\definecolor{5red}{HTML}{c9184a}

\definecolor{lightgreen}{RGB}{169, 245, 189}
\definecolor{lightred}{RGB}{222, 104, 116}

\newcommand{\method}{\textbf{AdvOF}}

\makeatletter
\begin{document}

\title{Disrupting Vision-Language Model-Driven Navigation Services via Adversarial Object Fusion}

\author{\and Chunlong Xie,
      \and Jialing He,~\textit{IEEE Member},
      \and Shangwei~Guo,~\textit{IEEE Member},
      \and Jiacheng~Wang,~\textit{IEEE Member},
      \and Shudong Zhang,
      \and Tianwei~Zhang,~\textit{IEEE Member},
      \and Tao~Xiang,~\textit{IEEE Senior Member},

\thanks{Chunlong Xie, Shangwei Guo, Jialing He, and Tao Xiang are with College of Computer Science, Chongqing University, Chongqing, China, 400044.}
\thanks{Jiacheng Wang is with the College of Computing and Data Science, Nanyang Technological University, Singapore 639798.}
\thanks{Shudong Zhang is with the School of Computer Science and Technology, Xidian University, Xi’an, China. 710071.}
\thanks{Tianwei Zhang is with College of Computing and Data Science, Nanyang Technological University, Singapore, 639798.}
\thanks{Shangwei Guo (swguo@cqu.edu.cn) and Jiacheng Wang (jiacheng.wang@ntu.edu.sg) are the corresponding authors.}
}

\IEEEtitleabstractindextext{
\begin{abstract}
We present Adversarial Object Fusion (\method{}), a novel attack framework targeting vision-and-language navigation (VLN) agents in service-oriented environments by generating adversarial 3D objects.
While foundational models like Large Language Models (LLMs) and Vision Language Models (VLMs) have enhanced service-oriented navigation systems through improved perception and decision-making, their integration introduces vulnerabilities in mission-critical service workflows. 
Existing adversarial attacks fail to address service computing contexts, where reliability and quality-of-service (QoS) are paramount.
We utilize \method{} to investigate and explore the impact of adversarial environments on the VLM-based perception module of VLN agents.
In particular, \method{} first precisely aggregates and aligns the victim object positions in both 2D and 3D space, defining and rendering adversarial objects. 
Then, we collaboratively optimize the adversarial object with regularization between the adversarial and victim object across physical properties and VLM perceptions. 
Through assigning importance weights to varying views, the optimization is processed stably and multi-viewedly by iterative fusions from local updates and justifications.
Our extensive evaluations demonstrate \method{} can effectively degrade agent performance under adversarial conditions while maintaining minimal interference with normal navigation tasks.
This work advances the understanding of service security in VLM-powered navigation systems, providing computational foundations for robust service composition in physical-world deployments.
\end{abstract}

}
\maketitle
\input{body/01_Introduction}

\input{body/02_Related_Work}

\input{body/03_Problem_Statement}

\input{body/04_Methodology}

\input{body/05_Experimental_Results}

\section{Conclusion}\label{sec:conclusion}
In this paper, we introduce a novel problem of adversarial objects targeting VLN agents integrated with foundation models. To tackle this problem, we propose \method{}, which leverages aligned object rendering, adversarial collaborative optimization and adversarial object fusion. \method{} enables precise localization of the victim object, facilitating alignment between 3D manipulations and 2D perceptions. Furthermore, \method{} can generate an effective adversarial object through collaborative optimization. Additionally, \method{} enhances attack performance by incorporating multiple views with importance weights. This proposed method presents a significant threat to the evolving capabilities of VLN agents empowered by foundation models.

\bibliographystyle{IEEEtran}
\bibliography{main}

\end{document}

%% file: body/01_Introduction.tex
\section{Introduction}
Service computing has advanced intelligent automation across cloud~\cite{wang2017efficient}, edge~\cite{kong2022edge}, and IoT platforms~\cite{lei2020groupchain}, with Vision-and-Language Navigation (VLN)~\cite{anderson2018vision} emerging as a critical component in real-world applications such as smart cities, autonomous delivery, and assistive robotics. VLN agents interpret human instructions and visually perceive environments to navigate unfamiliar environments.
Typical VLN approaches rely on representation learning~\cite{huang2019transferable}, reinforcement learning~\cite{wang2018look} and imitation learning~\cite{wang2019reinforced}, yet they often struggle with generalization in complex environments due to limited data and task-specific training. 
Recent advancements and applications in foundational models, particularly Large Language Models (LLMs)~\cite{brown2020language, achiam2023gpt, xiong2024search} and Vision Language Models (VLMs)~\cite{radford2021learning, liu2023visual, pasquadibisceglie2023jarvis, wang2024generative, zhang2024multi}, have addressed these limitations by significantly enhancing generalization capabilities of VLN agents~\cite{huang2023visual, gadre2023cows, dai2023think, shafiullahclip}.
By integrating foundation models into core VLN modules~\cite{xu2024survey, kawaharazuka2024real}, agents can better understand natural language instructions and perceive complex visual environments.
Specifically, LLMs facilitate high-level interaction and task planning~\cite{dai2023think, schumann2024velma}, while VLMs enhance low-level perception through improved feature extraction~\cite{radford2021learning} and scene recognition~\cite{kirillov2023segment}.
The integration of foundation models is gradually shaping a new deployment paradigm for VLN agents.

\IEEEpubidadjcol

\ExecuteMetaData[body/figure.tex]{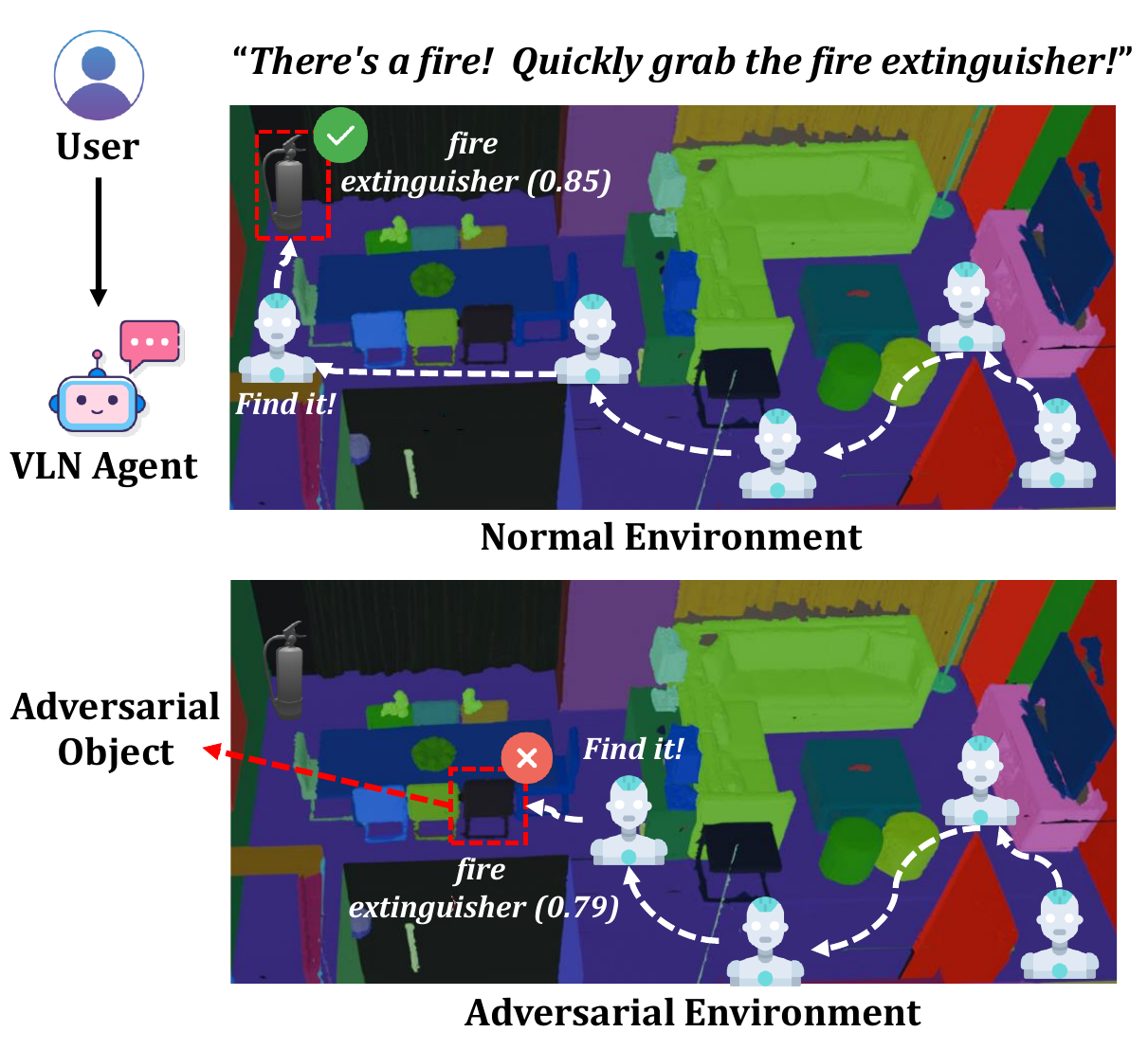}
However, this integration of foundation models could raise new security risks for VLN agents, particularly adversarial attacks~\cite{wang2021defending, liu2024exploring,lu2024poex,wang2024exploring,wen2024secure}.
Existing research has revealed robustness problems in foundation models through adversarial attacks. For example, attackers can manipulate LLMs to respond with unintended or harmful contents using adversarial suffix~\cite{zou2023universal, zhao2024accelerating} or jailbreak prompting~\cite{deng2023jailbreaker, mehrotra2024tree}. Similarly, adversarial perturbations added to input images can mislead VLMs into generating specified outputs~\cite{zhou2023advclip, zhao202evaluating}.
Building on these vulnerabilities, recent studies have identified robustness issues in the interaction and planning modules of LLM-powered agents~\cite{liu2024exploring, lu2024poex}. Such attacks typically exploit inherent flaws in LLMs by crafting malicious prompts that induce harmful agent behaviors. For instance, \cite{liu2024exploring} proposed a GCG-like~\cite{zou2023universal} optimization strategy to generate adversarial prompt suffix that misleads the planning module, evaluating the robustness of LLM-powered agents. \cite{lu2024poex} similarly optimized adversarial suffixes appended to harmful instructions, which jailbreak LLM-powered agents into producing executable harmful policies. 

Adversarial attacks targeting VLM-powered agents primarily focus on misleading the perception module. As illustrated in Fig.~\ref{fig:intro}, attackers can generate a 3D adversarial object in the environment, which misleads object perception (e.g., causing a ``chair" to be misclassified as a ``fire extinguisher"). Consequently, human instructions are incorrectly executed, leading to erratic agent behaviors. Thus, evaluating the robustness of VLM-powered VLN agents is critical. However, existing 2D adversarial attacks towards VLMs and 3D adversarial attacks on VLN agents face the following challenges for such evaluations. 
\revcircled{1} \textit{2D attacks suffer from attack misalignment}:
Traditional 2D adversarial attacks typically introduce pixel-wise perturbations or patch-based modifications to images. However, VLN agents operate in a 3D environment and perceive their surroundings through dynamic 2D images. 
This discrepancy leads to a fundamental misalignment: adversarial perturbations and patches designed for 2D perception may not align well with real-world 3D spatial constraints, limiting their effectiveness in practical scenarios.  
Therefore, adversarial objects must be constructed to align with physical properties while retaining attack efficacy.
\revcircled{2} \textit{2D attacks exhibit multi-view inefficiency}: 
2D attacks that inject 2D-space adversarial perturbations into images, render them effective only under specific, fixed views. 
Since VLN agents perceive objects in a 3D environment from diverse and unseen views, these adversarial perturbations effective in a static 2D image may become ineffective when the agent observes the same scene from a different angle or under varying lighting conditions.
Consequently, it is crucial to generate the adversarial object can consistently attack VLN agents, adapting to multiple views across varying navigation scenarios.
\revcircled{3} \textit{3D attacks face cross-modal inefficiency}: Traditional 3D attacks~\cite{liu2020spatiotemporal, huang2024towards} against VLN agents primarily target vision models like VGG~\cite{simonyan2014very}, R-CNN~\cite{he2017mask}, and ResNet~\cite{he2016deep}. These attacks, tailored for traditional visual models, face challenges when applied to VLM-powered agents due to the heterogeneity across different modalities~\cite{zhou2023advclip, zhao202evaluating}.

To address the aforementioned challenges, we propose a novel attack, Adversarial Object Fusion (\method{}), which generates adversarial objects capable of deceiving VLM-powered VLN agents. \method{} consists of three components, 
\textbf{1)} \textit{Aligned Object Rendering:} To solve the misalignment between 3D adversarial manipulation and 2D scene perception, we firstly aggregate object locations in 2D space using object detection and segmentation models. We then align these locations in 3D space to recognize isolated objects, enabling object-specific rendering from 3D space to 2D images;
\textbf{2) }\textit{Adversarial Collaborative Optimization:} To realize the cross-modal attack consistency, we design a collaborative optimization that captures universal adversarial features by constraining the similarity between visual and textual embeddings. Additionally, we combine an object-aware regularization term to preserve the physical plausibility of adversarial objects;
\textbf{3)} \textit{Adversarial Object Fusion:} To achieve reliable attack across varying and unseen views, we optimize the adversarial object across multiple views based on their importance weights and incorporate an iterative updating procedure to ensure stability in the optimization. This fusion guarantees the attack effectiveness across different VLN tasks and enhances the attack transferability across diverse VLN environments. 

Compared to existing 2D attacks (MF~\cite{zhao202evaluating} and AdvCLIP~\cite{zhou2023advclip}) and 3D attacks (ST~\cite{liu2020spatiotemporal} and TT3D~\cite{huang2024towards}), \method{} can achieve SOTA attack performance across four VLN agents (Vlmaps~\cite{huang2023visual}, Cows~\cite{gadre2023cows}, CF~\cite{shafiullahclip}, ORION~\cite{dai2023think}). Our main contributions are summarized as follows: 
\begin{itemize} 
    \item We formulate a new problem that generates adversarial objects towards VLN agents. 
    \item We develop a novel attack, \method{}, successfully generating the adversarial object that can multi-viewedly fool agents to percept it with the adversarial label. This adversarial object in the environment obstructs the execution of the user instruction.
    \item We conduct empirical validation of \method{}'s performance across multiple VLN agents and datasets, demonstrating its superiority in attacking effectiveness and multi-view robustness. 
\end{itemize}

%% file: body/02_Related_Work.tex
\section{Related Work}
\subsection{VLN Agents With Foundation Models}
Existing foundation models have demonstrated exceptional capabilities across several dimensions, including in-context learning~\cite{brown2020language}, reasoning~\cite{wei2022chain}, and multi-modal processing~\cite{radford2021learning}. VLN agents leveraging these models have similarly advanced, adopting novel implementation paradigms and achieving substantial performance gains. 

\nbf{Large Language Models for VLN Agents}. LLMs have shown promising capabilities in navigation, allowing VLN agents to follow interactive instructions and execute complex planning tasks. LEO~\cite{huang2024embodied} leveraged extensive knowledge from LLMs to excel in 3D perception, reasoning, and action tasks, training on large-scale 3D datasets and demonstrating remarkable proficiency across diverse real-world applications. LLM-Planner~\cite{song2023llm} introduced a hierarchical framework, where high-level plans composed of sub-goals are generated and subsequently refined into detailed actions by a low-level planner, enabling more adaptable and goal-oriented navigation strategies. NaviLLM~\cite{zheng2024towards} transformed embodied navigation tasks into structured generation problems via schema-based instructions, enabling generalization across diverse navigation scenarios with enhanced flexibility and consistency.

\ExecuteMetaData[body/figure.tex]{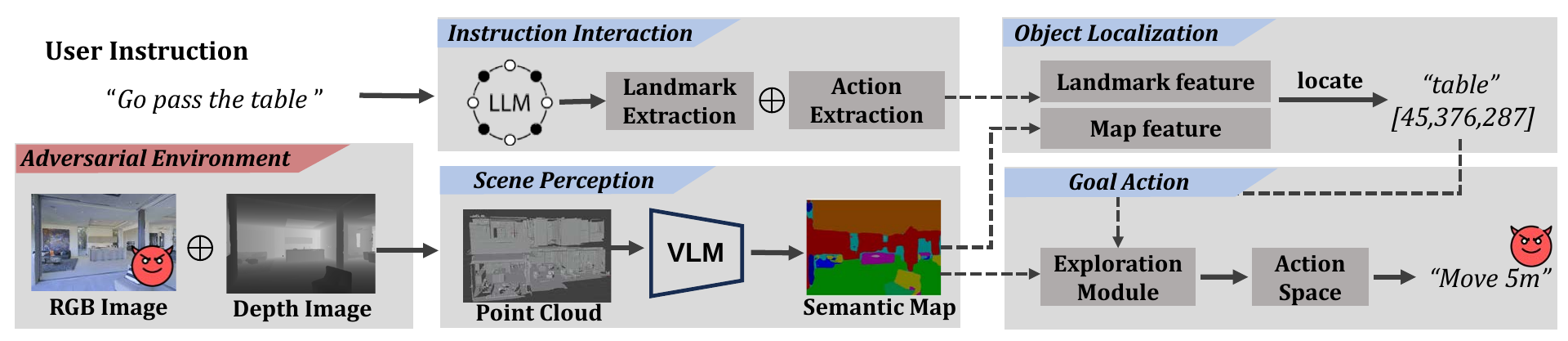}

\nbf{Vision Language Models for VLN agents}. 
With advancements in multi-modal representation learning, VLMs have shown remarkable performance in scene perception and map construction for VLN agents, achieving impressive zero-shot capability across diverse navigation tasks. 
Cows~\cite{gadre2023cows} investigated language-driven zero-shot object navigation, adapting open-vocabulary models to enable robots to locate objects specified through language without task-specific training.
CF~\cite{shafiullahclip} introduced an open-set, multi-modal 3D scene representation that integrates pixel-aligned features from pre-trained foundation models into 3D maps via SLAM, enabling zero-shot spatial reasoning across diverse queries.
ZSON~\cite{majumdar2022zson} utilized the vision encoder of  CLIP~\cite{radford2021learning} to encode target images, trained via a reinforcement learning framework. 
Vlmaps~\cite{huang2023visual} developed a spatial map representation that combines VLM features with 3D reconstructions, allowing robots to autonomously build maps from video feeds and support complex natural language navigation goals. 
ONION~\cite{dai2023think} leveraged multiple foundation models, enabling robots to navigate to personalized objects in unknown environments through user interaction.

\subsection{Adversarial Attack}
\nbf{2D Adversarial Attacks}. Adversarial attacks on foundation models have been extensively studied, revealing strategies to exploit model vulnerabilities. 
Textual adversarial attacks can mislead models into generating incorrect outputs or classifications \cite{yao2024survey, shen2023large}. Common techniques include token manipulation~\cite{li2020bert}, gradient-based optimization~\cite{zou2023universal}, and jailbreak prompting~\cite{deng2023jailbreaker}.
Visual adversarial attacks typically involve perturbing images to induce harmful content generation, often enhanced via proxy models~\cite{dong2023robust} or model ensembles~\cite{chenrethinking}. For example, AdvClip~\cite{zhou2023advclip} designed a universal adversarial patch targeting pre-trained VLMs, capable of compromising diverse downstream models. Similarly, MF~\cite{zhao202evaluating} introduced targeted adversarial examples against VLMs by leveraging transfer attacks with black-box queries.

In particular, some adversarial attacks have focused on LLM-powered agents. EIRAD~\cite{liu2024exploring} developed an embodied attack dataset for robustness evaluation, generated via the GCG algorithm~\cite{zou2023universal} with novel prompt suffix initialization. 
Similarly, NPS~\cite{wen2024secure} constructed an adversarial suffix attack targeting outdoor navigation agents, misleading them into navigating in incorrect directions. POEX~\cite{lu2024poex} designed a policy-executable red-teaming framework capable of injecting universal and transferable adversarial suffixes into planning modules, inducing embodied AI systems to execute harmful policies.

\nbf{3D Adversarial Attacks}. 3D adversarial attacks introduce physical perturbations to disrupt the inference of DNN-based models~\cite{wei2024physical}. Current 3D attacks are mainly implemented through gradient-based optimization~\cite{tsai2020robust, zheng2019pointcloud}, model-based generation~\cite{hamdi2020advpc, zhou2020lg}, and mesh-based transformations~\cite{suryanto2022dta, suryanto2023active}. 
For example, \cite{tsai2020robust} proposed robust adversarial objects against point cloud models in the physical world. 
\cite{zhou2020lg} designed an arbitrary-target attack framework using a label-guided adversarial network based on graph patch GAN architecture~\cite{shrivastava2017learning}. 
\cite{suryanto2023active} generated universal adversarial camouflage through neural texture rendering, incorporating stealthiness and naturalness constraints. 
ST~\cite{liu2020spatiotemporal} introduced 3D spatiotemporal perturbations that exploit temporal interaction history and spatial object properties, enhancing attack efficacy through a trajectory attention module.
TT3D~\cite{huang2024towards} produced transferable targeted 3D adversarial examples by reconstructing textured meshes and leveraging dual NeRF-space optimization to improve black-box transferability and visual naturalness.

However, these 2D and 3D adversarial attacks encounter limitations in generating adversarial objects for VLN agents. 2D adversarial attacks cannot perturb 3D space and exhibit limited effectiveness when executing attacks from multiple or unseen views. 3D adversarial attacks primarily target point cloud networks (e.g., PointNet~\cite{qi2017pointnet} and PointNet++~\cite{qi2017pointnet++}) and traditional detection models (e.g., Fast R-CNN~\cite{ren2016faster} and YOLO~\cite{redmon2016you}), which struggle to adapt to VLMs with cross-modal features and complex architectures.  Motivated by these problems, this paper focuses on designing a novel attack framework, capable of generating effective and multi-view adversarial objects targeting VLN agents.

%% file: body/03_Problem_Statement.tex
\section{Problem Statement}
This section describes the system model of VLN agents and outlines the threat model associated with an adversarial environment that contains adversarial objects. We then formalize the definition of adversarial objects targeting VLN agents.

\subsection{System Model}
We consider a VLN agent operating in a continuous environment~\cite{krantz2020beyond}, equipped with LLMs and VLMs. The agent executes primitive actions (e.g., move forward, turn left) to navigate toward a specified goal with a physical space, guided by natural language instructions. A VLN agent primarily comprises modules for scene perception, instruction interaction, object localization and goal action. The system architecture is illustrated in Fig.~\ref{fig:problem}.

\nbf{Instruction Interaction}. 
The VLN agent employs an interaction module based on LLMs to interpret user instructions, extracting object landmarks and associated actions. The user instruction is denoted as $L=\langle \bs{t}_0, \bs{t}_1, \dots, \bs{t}_L \rangle$, where $\bs{t}_i$ represents a single word token. Through querying the LLM, the agent identifies object landmarks $\langle o_1, o_2, \dots, o_m \rangle$ from $L$. For each landmark $o_i$, the LLM generates a corresponding action $\alpha_i$, guiding the agent to the target landmark. 

\nbf{Scene Perception}. The VLN agent constructs a point cloud using RGB and depth images to model its environment. Using these RGB-D images and the point cloud, the agent incrementally builds a semantic map that incorporates visual features extracted via VLMs. This map is represented as $\mathcal{M} \in \mathbb{R}^{H\times W \times C}$, where $H$ and $W$ denote the height and width of the map, and $C$ corresponds to the feature dimensionality of the VLM.

\nbf{Object Localization}. 
After extracting object landmarks from the interaction module, the agent locates their corresponding positions in the scene. It maintains a grounding label list and computes the grounding label feature using the VLM’s text encoder. The agent calculates similarity scores between each grounding label and the current semantic map $\mathcal{M}$, determining the target object location by selecting the highest-score map grid. 

\nbf{Goal Action}.
The agent executes goal-directed actions based on the identified object location and associated action within the predefined action space $\mathcal{A}$, including $rotate$, $move$, and $stop$. If the target landmark is not detected within the current camera view, the agent initiates scene exploration~\cite{yamauchi1997frontier}. Using the provided distance and angle information, the agent executes each action and, after each step, calculates its proximity to the object to determine whether it has reached the target location.
\ExecuteMetaData[body/figure.tex]{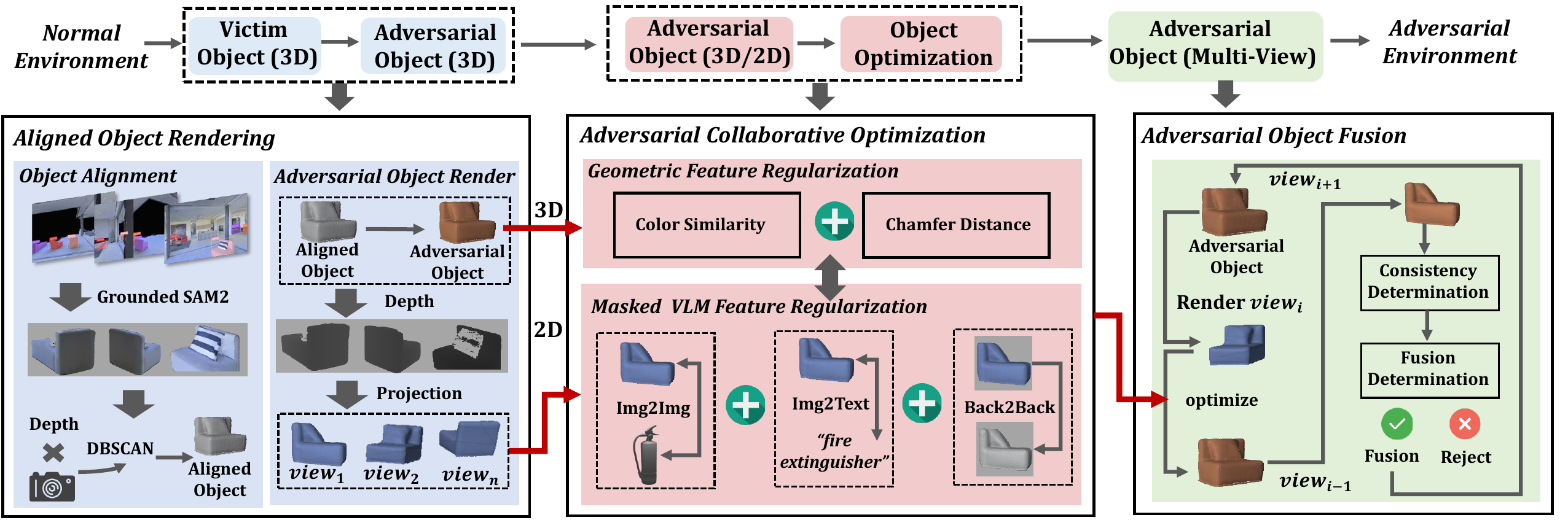}

\subsection{Threat Model}
\nbf{Attack Scenario}. 
While recent works have applied foundation models to enhance VLN agent performance under normal conditions~\cite{dai2023think, gadre2023cows}, studies reveal that these models remain highly sensitive to even minor input perturbations~\cite{shi2023large}. As illustrated in Fig.~\ref{fig:problem}, we examine an adversarial scenario targeting the scene perception module to disrupt instruction execution. Specifically, an adversary could introduce specially crafted objects to create an adversarial environment. When deployed in such an environment, the VLN agent misclassifies these adversarial objects into attacker-defined labels, leading to erroneous scene perception. Consequently, these errors propagate to the object localization and goal-oriented action modules, ultimately resulting in failed task execution.

\nbf{Attack Goals}. 
The attack targets scene perception modules powered by VLMs by introducing adversarial objects. The adversarial objects are designed to mislead the VLM into mislabeling them, distorting the semantic map and preventing accurate localization of target objects. As a result, the agent fails to reach the target object, thereby disrupting task completion. The following objectives outline a successful adversarial attack towards VLM-powered VLN agents:
\begin{itemize}
    \item \textit{Object-Specific Control}: The attack must target specific objects without affecting others in the environment. The adversary can select any object as a target, ensuring that the impact is contained within the targeted object area. 
    \item \textit{3D-2D Consistency}: The adversarial object, positioned in 3D space, should maintain its deceptive effect from different 2D views captured by the agent’s camera. This ensures the attack consistently impacts both the agent’s 3D understanding and 2D perception, leading to misinterpretation of the adversary object’s identity.
    \item \textit{Multi-View Robustness}: The adversarial object should deceive the agent from multiple views, ensuring adaptability to various user tasks and instructions.
\end{itemize}

\nbf{Adversary's Capabilities}. 
The adversary’s capabilities can be categorized as: \textbf{1)} \textit{Environment Manipulation}: The adversary has access to the system environment and can collect environmental data to support the attack. They can modify target objects (e.g., Painting~\cite{duan2020adversarial}, 3D printing~\cite{shahrubudin2019overview}) to create adversarial counterparts, aligning with standard environment modeling and scene perception processes. \textbf{2)} \textit{Agent Access}: Attacks can be carried out in either a white-box or black-box setting. In the white-box scenario, the adversary has access to the VLM of the VLN agent. In the black-box scenario, the adversary lacks knowledge of the agent’s VLM and uses proxy agents to craft the attack, transferring it to the target agent.

\subsection{Problem Formulation}
\begin{definition}[\textbf{Adversarial Object Against VLN Agents}]
Given a victim object $O$ with the label $T^{v}$ in 3D space, the goal is to generate a corresponding adversarial object  $O^{adv}$ such that the VLN agent perceives $O^{adv}$ with an adversarial label $T^{t}$ when observed from multiple views $\mathcal{V}$. 
The process of generating adversarial objects is formulated as an optimization problem:
\begin{align}
\arg\min_{O^{adv}} \mathcal{L}_{3D}(O^{adv},O) + \mathbb{E}_{v\in\mathcal{V}} \mathcal{L}_{2D}(O^{adv}_{v}, T^{t},\theta),
\end{align}
where $\theta$ presents the parameters of the VLM employed by the VLN agent; $\mathcal{L}_{3D}$ denotes a loss function that enforces similarity in physical property between $O$ and $O^{adv}$ in 3D space, ensuring that $O^{adv}$ resembles $O$ in appearance; $\mathcal{L}_{2D}$ denotes a loss function that encourages the VLM to perceive $O^{adv}_{v}$ as the adversarial label $T^{t}$ in each 2D view $v$.
\end{definition}

This optimization minimizes physical differences between $O$ and $O^{adv}$ while maximizing the misclassifications of $O^{adv}$ to $T^{t}$ across multiple views $\mathcal{V}$.

%% file: body/04_Methodology.tex
\section{Methodology}
\subsection{Motivation}
Designing an effective adversarial object against VLN agents within 3D environments presents unique challenges due to the integration with foundation models. We identify three main challenges:

\textbf{1) Aligning Adversarial Objects Across 3D and 2D Spaces.} VLN agents perceive their environment through 2D images while navigating in 3D space, resulting in a spatial misalignment between 3D adversarial manipulations and the agent’s 2D perception. Traditional 2D adversarial attacks on VLMs, often using pixel-level noise or patch-based perturbations~\cite{zhou2023advclip}, do not address this misalignment.

\textbf{2) Ensuring Attack Robustness Across Views and Tasks.} VLN agents frequently operate under varying user instructions, requiring adversarial objects to maintain effectiveness across different views and tasks. This adaptability is challenging due to the dynamic nature of foundation models, which process varying visual perspectives and instruction sets. 

\textbf{3) Maintaining Attack Effectiveness Among Different Modalities.} Compared to typical 3D adversarial attacks tailored for visual models, attacks targeting VLMs need to maintain the attack effectiveness between visual and textual modalities. 

\nbf{Pipeline}. To solve these challenges, we propose a novel adversarial attack framework \method{} (\textbf{Adv}ersarial \textbf{O}bject generation based on view \textbf{F}usion), targeting VLN agents powered by VLMs. Our method consists of three core components: Aligned Object Rendering, Adversarial Collaborative Optimization, and Adversarial Object Fusion. We provide the attacking pipeline in Fig.~\ref{fig:method}.

\subsection{Aligned Object Rendering}
Building on \textbf{Definition 1}, generating a successful adversarial object requires first identifying the victim object $O$ and subsequently constructing the initial adversarial object $O^{adv}$ within the environment. 

\nbf{Victim Object Alignment}. To detect the victim object, we leverage a VLM model for rapid open-set object detection, querying whether the victim object $O$ with the label $T^v$ is present in a given scene $S_i$:
\begin{equation}
    \mathcal{S}_v \leftarrow \{S_{i}|\mathcal{I}(VLM(S_i, T^v)=1, S_i\in\mathcal{S}\},
\end{equation}
where $\mathcal{S}$ denotes the set of all environment scenes. We then apply Grounding DINO~\cite{liu2023grounding} for precise object detection and SAM~\cite{kirillov2023segment} for object segmentation:
\begin{equation}\label{eq:seg}
    (MASK, SCORE) \leftarrow \{SAM(GD(S_{j},T^{v})), S_{j}\in\mathcal{S}_{v}\},
\end{equation}
where $MASK$ captures the spatial locations of victim objects, and $SCORE$ represents the segmentation confidence. To consolidate fragmented detections into discrete objects, we back-project $MASK$ into 3D space and cluster locations using DBSCAN~\cite{ester1996density}.  
\begin{align}\label{eq:align}
    MASK_{3D} \xleftarrow{back-project} MASK, \nonumber \\
    O \xleftarrow{select} DBSCAN(MASK_{3D}).  
\end{align}
A randomly selected cluster from the output is designated as the victim object $O$ in 3D space for the adversarial attack.

\nbf{Adversarial Object Render}. Since VLN agents with VLMs process only 2D observations, we create the adversarial object $O^{adv}$ by defining a 3D perturbation $\delta^{adv}$ and rendering adversarial 2D images based on this perturbation.
\begin{definition}[\textbf{3D Perturbation for Adversarial Object Generation}]
Given a victim object $O$, a 3D perturbation $\delta^{adv}$ is point cloud-wise noise with same shape as $O$. The adversarial object is generated by adding $\delta^{adv}$ to $O$: $O^{adv} \leftarrow O + \delta^{adv}$. The perturbation is constrained by an upper bound, $||\delta^{adv}||_{p} \leq \epsilon$. 
\end{definition}
We then render the adversarial object into 2D space. Assuming the environment camera has an intrinsic matrix $k_{int}$ and an extrinsic matrix $k_{ext}$, the rendering function $\mathcal{R}:O_{v} \leftarrow \mathcal{R}(O, v)$ projects the object on to a specific view $v$:
\begin{align}
    PC &= k_{ext}\cdot O, \nonumber\\ 
    O_{v} &= z^{-1}\cdot PC \cdot K_{int},
\end{align}
where $z$ is the depth value for view $v$, and $O_{v}$ is the rendered 2D projection. The optimization problem based on the rendering function is formulated as:
\begin{align}
    \arg\min_{\delta^{adv}} \mathcal{L}_{3D}(O^{adv}, O) &+ \mathbb{E}_{v\in\mathcal{V}} \mathcal{L}_{2D}(O_{v} + \delta^{adv}_{v}, T^{t}, \theta), \nonumber\\
    s.t. &\;\delta^{adv}_{v} \leftarrow \mathcal{R}(\delta^{adv}, v).
\end{align}
The 2D adversarial perturbation $\delta_{v}^{adv}$ is derived via the same rendering process $\mathcal{R}$. The resulting adversarial image in 2D space is defined as $O_{v}+\delta_{v}^{adv}$.

\ExecuteMetaData[body/figure.tex]{example}

\subsection{Adversarial Collaborative Optimization}
\label{sec:method_optimization}
This section explains the design of $\mathcal{L}_{3D}$ and $\mathcal{L}_{2D}$ in the above optimization problem, based on the adversarial collaborative optimization. This approach optimizes regularization across 3D and 2D spaces, capturing joint visual and textual modalities within the VLM. 

\nbf{Geometric Feature Regularization}. The purpose of $\mathcal{L}_{3D}$ is to maintain the physical property similarity between the adversarial object $O^{adv}$ and the victim object $O$. 
First, we constrain the RGB values to maintain color similarity:
\begin{equation}
   \mathcal{L}_{color}(O^{adv}, O) = ||O^{adv}-O||^{2}.
\end{equation}
For geometric shape similarity, we utilize Chamfer distance to align the 3D structure:
\begin{align}
    L_{CD}(O^{adv}, O) =&\sum_{x\in O^{adv}}\min_{y\in O}||x-y||^2 \nonumber\\
    &+ \sum_{y\in O}\min_{x\in O^{adv}}||y-x||^2.
\end{align}
Consequently, these components define the 3D loss:
\begin{align}
    \mathcal{L}_{3D}(O^{adv}, O)= \mathcal{L}_{color}(O^{adv}, O)  + \mathcal{L}_{CD}(O^{adv}, O)
\end{align}

\nbf{Masked VLM Feature Regularization}. The objective of $\mathcal{L}_{2D}$ is to mislead the VLM into misclassifying the adversarial object. Untargeted attacks involve preventing the VLM from identifying the adversarial object as the original victim object. A naive optimization approach minimizes the similarity between visual features of rendered adversarial and victim object images:
\begin{align}
    \mathcal{L}^{I2I}_{\theta}(O_v, \delta_v^{adv}) = \mathcal{F}(\theta(O_v+\delta_v^{adv}), \theta(O_v)), 
\end{align}
where $\mathcal{F}$ is the cosine similarity function and $\theta$ represents VLM parameters. However, whole-image optimization is overly sparse, failing to focus on adversarial object regions. Instead, we propose optimizing within the adversarial object mask. To extract region-specific features, we introduce a masked feature operator $\mathcal{M}$:
\begin{equation}
    \mathcal{M}^{\theta}(O_v, mask) = Flat(\theta(O_v)\odot mask),
\end{equation}
where $Flat$ flattens features and $\odot$ denotes element-wise multiplication. This reformulate $\mathcal{L}^{I2I}$:
\begin{align}
    \mathcal{L}^{I2I}_{\theta}(O_v, &\delta_v^{adv}, mask) =\nonumber\\&\mathcal{F}(\mathcal{M}^{\theta}(O_v+\delta_v^{adv}, mask),\mathcal{M}^{\theta}(O_v, mask)).
\end{align}
To enhance cross-modal attack efficacy, we regularize the alignment between adversarial visual features and victim object text embeddings $T^{v}$:
\begin{align}
    \mathcal{L}^{I2T}_{\theta}(O_v, &\delta_v^{adv}, mask, T^v) \nonumber\\&= \mathcal{F}(\mathcal{M}^{\theta}(O_v+\delta_v^{adv}, mask), \theta(T^{v}))
\end{align}
Empirically, constraining only the masked region disturbs background predictions (Fig.~\ref{fig:exampled}). To preserve background consistency, we add a regularization term:
\begin{align}
    \mathcal{L}^{B2B}_{\theta}(O_v, &\delta_v^{adv}, mask) \nonumber\\&= \mathcal{F}(\mathcal{M}^{\theta}(O_v+\delta_v^{adv}, \overline{mask}), \mathcal{M}^{\theta}(O_v, \overline{mask}))
\end{align}
Thus, the 2D loss function of the untargeted attack in view $v$ is then represented as:
\begin{align}
    \mathcal{L}^{UT}_{2D}&=\mathcal{L}^{I2I}_{\theta}(O_v, \delta_v^{adv}, mask)\nonumber\\&+ \alpha\cdot\mathcal{L}^{I2T}_{\theta}(O_v, \delta_v^{adv}, mask, T^v) \nonumber\\&- \beta\cdot\mathcal{L}^{B2B}_{\theta}(O_v, \delta_v^{adv}, mask)
\end{align}
where $\alpha$ and $\beta$ are balancing coefficients.

For targeted attacks, the goal is to misclasify $O$ as adversarial label $T^{t}$. To achieve this, we retain a target image $I^{t}$, with $mask^{t}$ for attack target $T^{t}$. Thus, the loss function of $\mathcal{L}_{\theta}^{I2I}$ for target attacks is revised as:
\begin{align}
    \mathcal{L}^{I2I}_{\theta}(O_v, &\delta_v^{adv}, mask, I^{t}, mask^{t}) =\nonumber\\&\mathcal{F}(\mathcal{M}^{\theta}(O_v+\delta_v^{adv}, mask),\mathcal{M}^{\theta}(I^{t}, mask^{t})).
\end{align}
Similarly, $\mathcal{L}_{\theta}^{I2I}$ aligns adversarial features with $T^t$:
\begin{align}
    \mathcal{L}^{I2T}_{\theta}(O_v, &\delta_v^{adv}, mask, T^t) \nonumber\\&= \mathcal{F}(\mathcal{M}^{\theta}(O_v+\delta_v^{adv}, mask), \theta(T^{t}))
\end{align}
Consequently, the 2D loss function of the targeted attack in view $v$ is represented as :
\begin{align}
    \mathcal{L}^{T}_{2D}&=-\mathcal{L}^{I2I}_{\theta}(O_v, \delta_v^{adv}, mask, I^{t}, mask^{t})\nonumber\\&- \alpha\cdot\mathcal{L}^{I2T}_{\theta}(O_v, \delta_v^{adv}, mask, T^t) \nonumber\\&- \beta\cdot\mathcal{L}^{B2B}_{\theta}(O_v, \delta_v^{adv}, mask)
\end{align}
\nbf{Exemplary Illustration}. Fig.~\ref{fig:example} demonstrates the optimization outcomes of targeted attacks using different regularization components. Notably, our proposed formulation $\mathcal{L}_{2D}^{T}$ effectively shifts adversarial object predictions from the victim label to the target label while preserving perceptual consistency in background regions.

\subsection{Adversarial Object Fusion}
In practical applications, adversarial objects must maintain effectiveness across diverse views. \textbf{Definition 1} formalizes this requirement as a uniform update across all views $\mathcal{V}$. However, views have varying levels of importance for the adversarial object. For example, incomplete or distant views are given lower weights, while complete or close-up views are assigned higher importance. Additionally, adjacent views frequently overlap when observing the same adversarial object. Therefore, 
a view-aware collaborative optimization strategy becomes critical.  

\ExecuteMetaData[body/algorithm.tex]{view}
\ExecuteMetaData[body/table.tex]{simulator}

We utilize $SCORE$ of the grounding model to represent the importance weight for each scene view. Besides, we also compute the pixel count $N$ of the adversarial object to indicate the importance weight $w_{v}$:
\begin{align}
    w_{v} = \frac{w}{\sum_{v\in V}{w}}, \text{where } w &= score_{v} + N_{v}/N 
\end{align}
This weighted approach modifies the optimization problem as follows:
\begin{align}\label{eq:vo}
    \arg\min_{\delta^{adv}} \mathcal{L}_{3D}(O^{adv}, O) + \mathbb{E}_{v\in\mathcal{V}} \; w_{v}\cdot \mathcal{L}_{2D}(O_v+\delta^{adv}_{v}, T^{t}, \theta).
\end{align}
To ensure stable multi-view updates, we propose an adversarial object fusion process that iteratively fuses the adversarial object across views, where a global perturbation $\delta^{adv}$ is refined through iterative updates from local perturbations for individual views.

The detailed algorithm of adversarial object fusion is presented in Algorithm~\ref{alg_viw}. In each iteration, the algorithm renders a 2D perturbation $\delta^{adv}_{v}$ from a 3D perturbation $\delta^{adv}$ (line 8). The 2D perturbation $\delta^{adv}_{v}$ is then optimized through the collaborative optimization method described in Sec 4.3 (Line 9). 
Next, consistency is evaluated with the prior view $v-1$ (Line 12). 
If the distance between the new perturbations $\delta^{adv}_{v-1}$ and prior local perturbations $\delta^{adv'}_{v-1}$ is larger than a consistency threshold $\mu_{1}$:
\begin{equation}
MSE(\delta^{adv'}_{v-1}, \delta^{adv}_{v-1}))\geq \mu_{1}, 
\end{equation}
the local update is rejected, and the importance weight is reduced before reattempting the collaborative optimization. Following this, a fusion determination step assesses whether the local perception aligns with the previous view's fusion (Line 14). If the discrepancy of these perceptions exceeds a fusion threshold $\mu_2$:
\begin{equation}
|\mathcal{L}_{2d}(\delta_{v-1}^{adv'})-\mathcal{L}_{2d}(\delta_{v-1}^{adv})|\geq \mu_{2},
\end{equation}
this local update is discarded, and the perturbation bound is reduced before re-rendering. If the view fails to meet these conditions within $Max$ iterations, the fusion is rejected for that view. By iterating through all views, the algorithm robustly fuses local perturbations into a cohesive adversarial object across views.

%% file: body/05_Experimental_Results.tex
\ExecuteMetaData[body/figure.tex]{simulator}
\section{Experiments}
\ExecuteMetaData[body/table.tex]{instruction}
\ExecuteMetaData[body/table.tex]{overall}
\ExecuteMetaData[body/figure.tex]{noise}
In this section, we first present the primary attack results against various VLN agents compared to 2D and 3D baseline attacks (Section~\ref{sec:overall}). Next, we demonstrate the transferability of \method{} across diverse image encoders, scene datasets, and model architectures (Section~\ref{sec:transfer}). We then assess the robustness of the attack under potential defensive mechanisms (Section~\ref{sec:defense}). Finally, we perform parameter analysis to evaluate the impact of critical modules and hyperparameters (Section~\ref{sec:parameter}).
\subsection{Environmental Setup}

\nbf{VLN Agents}. We adopted four typical navigation agents using foundation models. 1) Vlmaps~\cite{huang2023visual}: integrates language grounding with visual observations through a spatial map that fuses pre-trained visual-language features~\cite{li2022languagedriven} with a 3D reconstruction of the physical world;
2) Cow~\cite{gadre2023cows}: introduces the PASTURE benchmark for language-driven zero-shot object navigation, adapting zero-shot models to a VLN task;
3) CF~\cite{shafiullahclip}: constructs an implicit scene model based on clip~\cite{radford2021learning} to strengthen instance identification and semantic segmentation; 
4) ORION~\cite{dai2023think}: proposes Zero-shot Interactive Personalized Object Navigation, requiring robots to navigate to personalized goal objects while engaging in conversations with users.

\ExecuteMetaData[body/table.tex]{labelacc}

\nbf{Simulation Datasets and Environments}. For each navigation agent, we followed the original simulator and dataset settings as provided in the official repository. Table~\ref{tab:simulator} details the simulation environments, with dataset visualizations presented in Fig.~\ref{fig:simulator}. 

The simulator adopted in these navigation agents is Habitat~\cite{savva2019habitat}, enabling highly efficient photorealistic 3D simulation. The datasets used are Matterport3D (MP3D)~\cite{chang2017matterport3d} and Habitat-Matterport 3D (HP3D)~\cite{ramakrishnan2021habitatmatterport}. MP3D is an RGB-D dataset with 90 building-scale scenes, and HM3D is a large-scale dataset of 1,000 building-scale scenes. 

For each navigation agent and dataset, we randomly selected 20 scenes to construct the validation dataset. In each scene, we randomly selected 10 objects as victim objects. For each object, we randomly selected 10 different instructions. The instructions collected in different navigation agents are displayed in Table~\ref{tab:instruction}.

\nbf{Baselines and Metrics}. We adopted two SOTA adversarial attacks to VLMs and two 3D adversarial attacks to traditional object detection models as the baselines. 1) MF~\cite{zhao202evaluating}: evaluates the robustness of open-source large VLMs
under black-box conditions, where adversaries aim to mislead the model into returning the targeted responses; 2) AdvCLIP~\cite{zhou2023advclip}: generates downstream-agnostic adversarial examples based on cross-modal pre-trained encoders. 3) ST~\cite{liu2020spatiotemporal}: studies adversarial attacks on embodied agents using spatiotemporal perturbations in dynamic environments. 4) TT3D~\cite{huang2024towards}: creates transferable targeted 3D adversarial examples using NeRF-based optimization for improved black-box transferability.

We evaluate task and attack performance using standard object navigation metrics and adversarial attack measures:
1) \textit{Success Rate(SR)}: the fraction of episodes where the agent executes STOP action within 1.0m of the target object;
2) \textit{Success weighted by inverse path length (SPL)}: Success weighted by the oracle shortest path length and normalized by the actual path length. This metric points to the success efficiency of the agent;
3) \textit{Key point accuracy (KPA)}: The KPA metric measures the percentage of correct decisions made at each sub-goal;
4) \textit{Acc}: The prediction accuracy of the scene perception module;
5) \textit{Asr}: The attack success rate in misleading the perception module’s predictions. 

\nbf{Attack Implementations}. In the implementation of the attack, the VLM model utilized for rapid open-set object detection is LLaVA~\cite{liu2023visual}. The box threshold and the text threshold for the grounding model are set to $0.40$. We set the balance coefficients $\alpha = 0.5$ and $\beta=0.01$, The upper bound of adversarial perturbation is set to $\epsilon = 32/255$. The optimizer used is Adam, with the optimization iterations set to $200$. The parameters $\mu_1$ and $\mu_2$ in the adversarial object fusion are set to $0.01$ and $0.05$, respectively. After attacking the victim object, we replace the RGB-D data associated with the victim object using the perturbed data and regenerate the semantic map. The attack performance is then evaluated in the new environment scene using the collected validation dataset. For the two attacks targeting VLMs, MF and AdvCLIP, we select RGB images containing the victim object to be attacked. For the two 3D attacks, ST and TT3D, we first implement the proposed process of aligned object rendering and then proceed with the attack as outlined in the original paper.

\ExecuteMetaData[body/figure.tex]{view}

\subsection{Overall Evaluation}\label{sec:overall}
\nbf{Effectiveness}. To comprehensively evaluate the impact of adversarial objects, we construct a new validation dataset without adversarial objects. We present the overall attack results (Table~\ref{tab:overall}),  with attack objects (denoted as ``attacked") and without attack objects (denoted as ``normal") in a white-box scenario against various navigation agents, including non-interactive agents (Vlmaps~\cite{huang2023visual}, Cow~\cite{gadre2023cows}, CF~\cite{shafiullahclip}) and interactive agents (ORION~\cite{dai2023think}). Examples of the adversarial objects are shown in Fig.~\ref{fig:noise}.
\ExecuteMetaData[body/table.tex]{sizetransfer}
\ExecuteMetaData[body/table.tex]{datatransfer}

In 2D adversarial attacks, the attack surface typically consists of RGB images. As these attacks are not optimized for 3D environments, their impact on navigation agents remains limited. MF applies global noise affecting the entire image, but its effect on specific adversarial targets remains constrained. This limitation manifests in minimal changes to the agent's perceptual confidence, exemplified in Fig.~\ref{fig:noise_mf}, where the `chair' confidence score merely decreases from 0.44 to 0.41. However, despite its limited efficacy on adversarial objects, MF significantly degrades navigation performance on normal datasets. By contrast, AdvCLIP utilizes patch-based perturbations independently optimized per view. As demonstrated in Fig.~\ref{fig:noise_advclip}, AdvCLIP achieves greater disruptive efficacy, reducing the `chair' confidence score from 0.44 to 0.13 compared to MF's modest decrease (0.44 to 0.41). Nevertheless, since AdvCLIP optimizes patches for individual views, this approach introduces cross-view inconsistencies. While effective for previously observed viewpoints, AdvCLIP cannot generalize to novel perspectives. By concentrating perturbations on victim object regions while minimizing effects on other elements, AdvCLIP better preserves navigation performance on normal datasets.

In 3D adversarial attacks, the attack surface operates in 3D space. While these methods consider the relationship between 3D spatial attacks and 2D images, they are not specifically optimized for VLMs, resulting in reduced effectiveness on 2D images. ST (Fig.~\ref{fig:noise_st}) optimizes adversarial objects using trajectory historis, achieving success in scenarios with similar trajectories. However, ST demonstrates poor attack efficacy when following divergent instructions that produce anomalous trajectories. TT3D (Fig.~\ref{fig:noise_tt3d}) focuses more on victim objects themselves, yielding stronger attack performance. Nevertheless, its efficacy is constrained by the complexity of VLMs employed in perception modules. Additionally, both attacks exhibit non-negligible impacts on the surroundings of adversarial objects, inducing perceptual interference on other objects in normal datasets.

In contrast, \method{} achieves a well-balanced attacking, effectively disrupting navigation performance on attacked datasets while preserving performance on normal datasets. This balance is evident from Table~\ref{tab:overall}, where \method{} outperforms both MF and AdvCLIP, demonstrating superior adversarial efficacy without compromising normal task performance. More importantly, Fig.~\ref{fig:noise}-(d) demonstrates that \method{} can precisely manipulate the perception of the adversarial object (chair:0.44 $\rightarrow$ table:0.81).

\nbf{Perception Performance}. 
\ExecuteMetaData[body/table.tex]{modeltransfer}
\ExecuteMetaData[body/table.tex]{physicalnoise}
To provide deeper insight into the attack effectiveness and validate the root cause of navigation failures, we conduct a targeted evaluation of the perception module's robustness. we sampled 20 images from diverse new perspectives for each victim object in the validation datasets and evaluated the performance of VLMs on these images. The results are presented in Table~\ref{tab:label}. Results consistent with those in Table~\ref{tab:overall} can be observed. The two 2D attack methods fail to mislead the VLM's predictions from new perspectives. In contrast, the two 3D attack methods demonstrate some attack efficacy, but their performance is inconsistent and susceptible to environmental interference. Conversely, AdvOF achieves stable attack efficacy from new perspectives to a significant extent.

\nbf{Multi-View Robustness}. We validated the multi-view robustness in Fig.~\ref{fig:view}. Clearly, \method{} can successfully generate the adversarial object from a long-range view (view1) to a close-up view (view4). Additionally, \method{} can precisely control the effectiveness of adversarial object within its own region, without affecting the perception of other objects in the environment. This object-specific control, combined with multi-view robustness, provides a more powerful attack in VLN tasks compared to baseline attacks.

\subsection{Transferable Evaluation}\label{sec:transfer}
In this section, we further examine the adversarial transferability of \method{} across different scenarios.

\ExecuteMetaData[body/figure.tex]{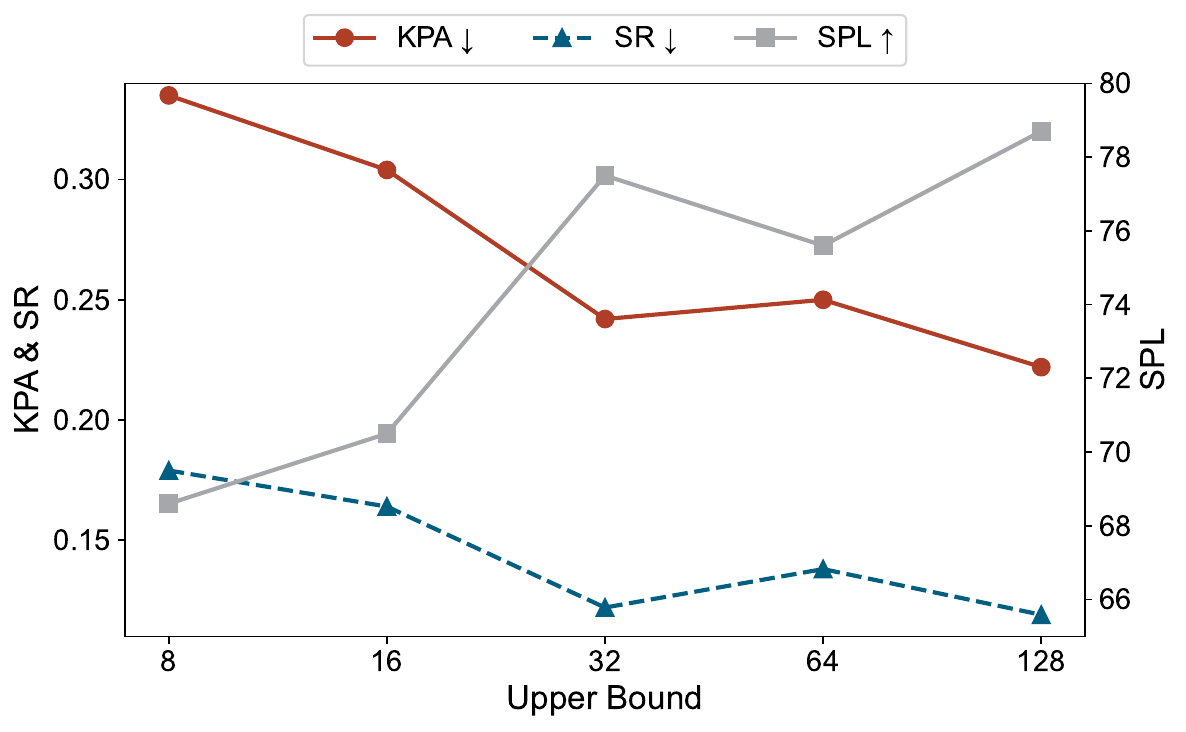}
\ExecuteMetaData[body/figure.tex]{resilience}
\ExecuteMetaData[body/figure.tex]{goal}

\nbf{Different Image Encoders}. We utilized different image encoder backbones to evaluate the transferability of adversarial attacks across various model architectures. For the Lseg model, we selected ViT-L/16 and ResNet101 as the image encoder backbones, while for the Clip model, we used ViT-B/32 and ViT-L/16. As shown in Table~\ref{tab:sizetransfer}, where in each category, the first row represents the performance of attacks generated and tested on the same backbone, and the second row represents the performance when transferred to a different backbone, with the decline rate indicated in red, our results demonstrate strong transferability of adversarial attacks across different image encoders. Specifically, the largest observed decline in performance is only 25.0\%. Furthermore, in both the Lseg and Clip models, the adversarial attacks transfer effectively between different encoders, exhibiting minimal degradation in performance.

\nbf{Different Scene Datasets}. We constructed different scene datasets featuring the same victim objects to evaluate the transferability of \method{} across validation datasets. The corresponding results are shown in Table~\ref{tab:datatransfer}, where \method{} demonstrates excellent transferability, with almost no degradation in attack performance across datasets (the largest decline is just 7.9\%). In fact, different scene datasets for the same adversarial object simply represent different views. Thus, \method{}, with its multi-view robustness, confirms its effectiveness in such transferable settings.

\nbf{Different Model Architectures}. Furthermore, we evaluated the transferability of our method across different navigation agents that utilize various VLMs. This evaluation aligns with the practical consideration of a black-box scenario. The corresponding results are presented in Table~\ref{tab:modeltransfer}. In each category of the table, the first row shows the performance of attacks generated and tested on the same VLM. The second row displays the best result from the baseline methods for that VLM. The third row indicates the performance of attacks when transferred to a different VLM. Compared to the white-box attack scenario, the effectiveness of the attack decreases significantly in the black-box setting. However, our method still achieves performance comparable to that of baseline attacks in a white-box scenario, demonstrating its transferable capability in black-box conditions.

\subsection{Possible Defense}\label{sec:defense}
In practice, the perception of an adversarial object is influenced by environmental factors such as camera pose, light intensity, distance, and other noise. To evaluate the robustness of \method{} against physical noise defenses, we utilized an image processing toolkit\footnote{https://github.com/aleju/imgaug} to perturb the adversarial objects. Specifically, we applied five image processing techniques: shearing, scaling, Gaussian noise, brightness enhancement, and brightness reduction. The parameters for these transformations were configured as follows: shear(-16, 16), scaling(0.8), Gaussian noise(0.1), brightness enhancement(1.5), and brightness reduction(0.5). The results are summarized in Table~\ref{tab:physicalnoise} and sample perturbed images are illustrated in Fig.~\ref{fig:resilience}. It is evident that \method{} exhibits remarkable robustness across various types of image noise, providing enhanced adaptability in real-world scenarios and under diverse environmental conditions. Moreover, the noise types of shearing and Gaussian noise exert a greater influence on adversarial objects, whereas brightness changes have a lesser impact.

\subsection{Parameter Analysis}\label{sec:parameter}
\ExecuteMetaData[body/table.tex]{ablation}
\nbf{Upper Bound Selection}. We study the effect of different adversarial upper bounds on the attack performance of \method{}. The corresponding results are shown in Fig.~\ref{fig:epsilon}. The attack performance gradually increases as the upper bound increases, but satisfactory performance can generally be achieved with a low bound ($\epsilon=32$).

\nbf{Adversarial Object Selection}. For a VLN task, a navigation instruction often involves multiple object goals. Therefore, we investigate how adversarial objects affect the navigation of other object goals. As shown in Fig.~\ref{fig:goal} (sub-goal@i, where $i$ represents the goal order in the navigation instruction or trajectory), we attacked victim objects with different position orders. When the victim object positioned at the front is attacked, it significantly impacts the navigation performance of subsequent goals. This occurs because the agent moves to a substantially different area, which is less likely to contain other goals. In contrast, when the victim object positioned last is attacked, the impact on navigation performance is smaller, as the adversarial goal only disrupts itself, leaving the preceding goals unaffected.

\nbf{Ablation Study}. 
We examined the contribution of each component in the proposed attack framework. The corresponding results are shown in Table~\ref{tab:ablation}. $Alignment$ refers to the aligned object rendering, which helps locate the 2D and 3D positions of the victim object. We test \method{} by directly adopting the 2D positions recognized by SAM. The results demonstrate that these identified positions are disordered and fail to reveal the true 3D positions of the victim object. $\mathcal{L}_{2d}$ defines the objective function in 2D space, which directly influences the scene perception module of the VLN agent and plays a crucial role in generating the adversarial object. $\mathcal{L}_{3d}$ constrains the perturbation in 3D space, primarily helping to optimize the adversarial object with respect to physical properties of the victim object. Fusion refers to the adversarial object fusion, which helps integrate the perturbation based on views with different importance weights. It contributes to the stable convergence of the optimization process and the generation of the optimized adversarial object.